\documentclass[fleqn]{stacs_proc} 
\usepackage{amssymb}
\usepackage{algorithm}
\usepackage{algorithmic}

\begin{document}
\title[Design by Measure and Conquer: A Faster Algorithm for Dominating Set]{Design by Measure and Conquer \\ A Faster Exact Algorithm for Dominating Set}
\author{Johan M. M. van Rooij}{Johan M. M. van Rooij}
\author{Hans L. Bodlaender}{Hans L. Bodlaender}
\address{
Institute of Information and Computing Sciences, Utrecht University, \newline
P.O.Box 80.089, 3508 TB Utrecht, The Netherlands
}
\email{jmmrooij@cs.uu.nl}
\email[Hans L. Bodlaender]{hansb@cs.uu.nl}
\urladdr{http://www.cs.uu.nl}

\keywords{exact algorithms, exponential time algorithms, branch and reduce, measure and conquer, dominating set, computer aided algorithm design}
\subjclass{
F.2.2. [Analysis of Algorithms and Problem Complexity]: Non-numerical Algorithms and Problems---computations on discrete structures;  
G.2.2. [Discrete Mathematics]: Graph Theory---graph algorithms;
I.2.2. [Artificial Intelligence]: Automatic Programming---automatic analysis of algorithms}

\begin{abstract}
The {\em measure and conquer} approach has proven to be a powerful tool to
{\em analyse} exact algorithms for combinatorial problems, like {\sc Dominating Set} and
{\sc Independent Set}. In this paper, we propose to use measure and conquer
also as a tool in the {\em design} of algorithms. 
In an iterative process, we can obtain a series of {\em branch and reduce} algorithms.
A mathematical analysis of an algorithm in the series with measure and conquer
results in a quasiconvex programming problem.
The solution by computer
to this problem not only gives a bound on the running time,
but also can give a new reduction rule, thus giving a new,
possibly faster algorithm.
This makes {\em design by measure and conquer} a form of {\em computer aided algorithm design}.

When we apply the methodology to a {\sc Set Cover} modelling of
the {\sc Dominating Set} problem, we obtain the
currently fastest known exact algorithms for {\sc Dominating Set}: an algorithm
that uses $O(1.5134^n)$ time and polynomial space, and an algorithm
that uses $O(1.5063^n)$ time.
\end{abstract} 

\maketitle

\stacsheading{2008}{657-668}{Bordeaux}
\firstpageno{657}

\section{Introduction}
The design of fast exponential time algorithms for finding exact solutions to
NP-hard problems such as {\sc Independent Set} and {\sc Dominating Set}
has been a topic for research for over 30 years, see e.g., the results
on {\sc Independent Set} in the 1970s by Tarjan and Trojanowski 
\cite{Tarjan72a,TarjanT77}. A number of different techniques have been used
for these and other exponential time algorithms \cite{FominGK05,Woeginger03,Woeginger04}.

An important paradigm for the design of exact algorithms is {\em branch and reduce},
pioneered in 1960 by Davis and Putnam \cite{DavisP60}. Typically, in a branch
and reduce algorithm, a collection of reduction rules
and branching rules are given.
Each reduction rule simplifies the instance to an equivalent, simpler
instance. If no rule applies, the branching rules generate a collection of two or more instances,
on which the algorithm recurses.

An important recent development in the analysis of branch and reduce algorithms is
{\em measure and conquer}, which has been introduced by Fomin, Grandoni and Kratsch \cite{FominGK05a}.
The measure and conquer approach helps to obtain good upper bounds on the running time
of branch and reduce algorithms, often improving upon the currently best known bounds for
exact algorithms. It has been used successfully on 
{\sc Dominating Set} \cite{FominGK05a}, {\sc Independent Set} \cite{FominGK06},
{\sc Dominating Clique} \cite{KratschL06}, the number of minimal dominating sets
\cite{FominGPS05}, {\sc Connected Dominating Set} \cite{FominGK06a}, {\sc Minimum Independent
Dominating Set} \cite{GaspersL06}, and possibly others.

In this paper, we show that the measure and conquer approach can not only be used
for the {\em analysis} of exact algorithms, but also for the {\em design} of such algorithms.
More specifically, measure and conquer uses a non-standard size measure for instances.
This measure is based on weight vectors, which are computed by solving a quasiconvex programming problem.
Analysis of the solution of this quasiconvex program yields not only an upper bound
to the running time of the algorithm, but also shows where we should improve
the algorithm. This can lead to a new rule, which we add to the branch and
reduce algorithm. 

We apply this {\em design by measure and conquer} methodology to a {\sc Set Cover}
modelling of the {\sc Dominating Set} problem, identical to the setting in which
measure and conquer was first introduced.
If we start with the trivial algorithm, then, we can obtain in a number of steps the original
algorithm of Fomin et al.~\cite{FominGK05a}, but with additional steps, we obtain a faster
algorithm, using $O(1.5134^n)$ time and polynomial space, with a variant that uses exponential
memory and $O(1.5063^n)$ time. We also show that at this point we cannot improve this
measure and conquer computed running time, unless we choose a different measure
or change the branching rules.

While for several classic combinatorial problems, the first non-trivial exact algorithms date many years
ago, for the {\sc Dominating Set} problem, the first algorithms with running time 
$O^\ast(c^n)$ with $c<2$ are from 2004, with three independent papers: by Fomin et
al.~\cite{FominKW04}, by Randerath and Schiermeyer \cite{RanderathS05}, and by
Grandoni \cite{Grandoni06}. The so far fastest algorithm for {\sc Dominating Set} was
found in 2005 by Fomin, Grandoni, and Kratsch \cite{FominGK05a}: this algorithm uses 
$O(1.5260^n)$ time and polynomial space, or $O(1.5137^n)$ time and exponential space.

\section{Preliminaries} \label{sec:definitions}
Given a collection of non-empty sets $\mathcal{S}$,
a \emph{set cover} of $\mathcal{S}$ is a subset
$\mathcal{C} \subseteq \mathcal{S}$ such that every element in any of the sets in $\mathcal{S}$ occurs in some set in $\mathcal{C}$.
In the {\sc Set Cover} problem we are given a collection $\mathcal{S}$
and are asked to compute a set cover of minimum cardinality.

The universe $\mathcal{U_S}$ of a {\sc Set Cover}
problem instance is the set of all elements in any set in $\mathcal{S}$; $\mathcal{U_S} = \bigcup_{S \in \mathcal{S}} S$.
The frequency $f(e)$ of an element $e \in \mathcal{U_S}$ is the number of sets in $\mathcal{S}$ in which this element occurs.
Let $\mathcal{S}(e) = \{ S \in \mathcal{S} ~|~ e \in S \}$ be the set of sets in $\mathcal{S}$ in which the element $e$ occurs.
We define a connected component $\mathcal{C}$ in a {\sc Set Cover} problem instance $\mathcal{S}$ in a natural way:
a minimal non-empty subset $\mathcal{C}\subseteq\mathcal{S}$ for which all elements in the sets in $\mathcal{C}$
occur only in sets contained in $\mathcal{C}$.
The dimension $d_\mathcal{S}$ of a {\sc Set Cover} problem instance is the sum
of the number of sets in $\mathcal{S}$ and the number of elements in $\mathcal{U_S}$; $d_\mathcal{S} = |\mathcal{S}| + |\mathcal{U_S}|$.

Let $G=(V,E)$ be an undirected graph.
A subset $D \subseteq V$ of nodes is called a \emph{dominating set}
if every node $v \in V$ is either in $D$ or adjacent to some node in $D$.
The {\sc Dominating Set} problem is to compute for a given graph $G$
a dominating set of minimum cardinality.

We can reduce the minimum dominating set problem to the
{\sc Set Cover} problem by introducing a set for each node of $G$ which contains
the node itself and its neighbours; $\mathcal{S} := \{ N[v] \;|\; v \in V \}$.
Thus we can solve a {\sc Dominating Set} problem on a graph of $n$ nodes by a
minimum set cover algorithm running on an instance of dimension $d = 2n$.

Both problems are long known to be NP-complete \cite{GareyJ79,Karp72}, which motivates
the search for fast exponential time algorithms.

\section{A Faster Exact Algorithm for Dominating Set} \label{sec:m&calgorithm}
In this section, we give our new algorithm for {\sc Dominating Set}. The
algorithm is an improvement to the algorithm by Fomin et
al.~\cite{FominGK05a}; it is obtained from this algorithm by adding some
additional reduction rules. These rules were derived using the design by
measure and conquer approach, see Section~\ref{section:design}. After
introducing our algorithm, we recall the necessary background of the measure
and conquer method \cite{FominGK05a}. 

\subsection{The Algorithm}
Our algorithm for the {\sc Dominating Set} problem uses the {\sc Set Cover}
modelling of {\sc Dominating Set} shown in Section \ref{sec:definitions}.
It is a branch and reduce algorithm on this modelling consisting of four simple reduction rules,
one base case for the recursion and a branching rule. See Algorithm \ref{algorithm}.

For a given problem instance we first apply the following reduction rules:
\begin{enumerate}
\item {\em Base case}. \label{rule:matching}
	  If all sets in the instance are of size at most two then finding
	  a minimum set cover is equivalent to finding a maximum matching in a graph.
	  Introduce a node for each element and an edge for each set of size two.
	  Now the maximum matching joined with some sets containing the unmatched nodes
	  form a minimum set cover. This matching can be computed in polynomial time \cite{Edmonds65}.
\item {\em Splitting connected components}. \label{rule:component}
		If the instance contains multiple connected components, compute the minimum set cover in each connected component separately.
\item {\em Subset rule}. \label{rule:subset}
	  If the instance contains sets $S_1$, $S_2$ with $S_1 \subseteq S_2$,
	  then we remove $S_1$ from the instance. Namely, in each minimum set
	  cover that contains $S_1$, we can replace $S_1$ by $S_2$ and obtain
	  a minimum set cover without $S_2$.
\item {\em Subsumption rule}. \label{rule:subsumption}
	  If the set of sets $S(e')$ in which an element $e'$ occurs is a subset of the set of sets $S(e)$
	  in which another element $e$ occurs, we remove the element $e$.
	  For any set cover, covering $e'$ also covers $e$.
\item {\em Unique element or singleton rule}.
	  If any set of size one remains in the instance after application of the previous rules, 
	  we add this set to the set cover. For the element in this set must occur uniquely in this set,
	  otherwise it would have been a subset of another set and have been removed by rule \ref{rule:subset}.
\item {\em Avoiding unnecessary branchings based on frequency two elements}. \label{rule:special}
	  For any set $S$ in the problem instance let $r_2$ be the number of frequency two elements it contains.
	  Let $m$ be the number of elements in the union of sets containing the other occurrences of these
	  frequency two elements, excluding any element already in $S$.
	  If for any set $S$: $m < r_2$ then include $S$ in the set cover. \\
	  This rule is correct since if we would branch on $S$ and include it in the set cover we would
	  cover $|S|$ elements with one set. If the set cover does not use $S$,
	  it must use 
	  all sets containing the other occurrence of the frequency two elements,
	  since they have become unique elements now. Notice that by Rule \ref{rule:subsumption}
	  all other occurrences of the frequency two elements must be in different sets and
	  thus we would cover $|S| + m$ elements with $r_2$ sets.
	  So if $m < r_2$ the first case can be preferred over the second: we can just add $S$ to the
	  cover and have $r_2 - 1 \leq m$ sets left to cover at least this much elements.
\end{enumerate}
For the branching rule, we select a set of maximum cardinality and
create two subproblems by either including it in the minimum set cover and removing all
newly covered elements from the problem instance or removing it.

\begin{algorithm}
\begin{algorithmic}
\STATE MSC($\mathcal{S}$) = \{
\IF{$\max\{|S| \; | \; S \in \mathcal{S} \} \leq 2$}
	\RETURN minimum set cover of $\mathcal{S}$ by computing a matching
\ELSIF{$\exists \mathcal{C}\subseteq\mathcal{S} : \mathcal{C}\not\in\{\emptyset,\mathcal{S}\}, \{S(e) | e \in S, S \in \mathcal{C} \} = \mathcal{C}$}
	\RETURN MSC($\mathcal{C}$) + MSC($\mathcal{S}\backslash\mathcal{C}$)
\ELSIF{$\exists S,S' \in \mathcal{S} : S \not= S', \; S \subseteq S'$}
	\RETURN MSC($\mathcal{S}\backslash\{S\}$)
\ELSIF{$\exists e,e' \in \mathcal{U_S} : e \not= e', \; \mathcal{S}(e) \supseteq \mathcal{S}(e')$}
	\STATE $\mathcal{S'} = \mathrm{MSC}(\{S \backslash \{e\} | S \in \mathcal{S} \}) $
	\RETURN $\{S \cup \{e\} | S \in \mathcal{S'}, S \cup \{e\} \in \mathcal{S}\} \cup \{S | S \in \mathcal{S'}, S \cup \{e\} \not\in \mathcal{S} \}$
\ELSIF{$\exists \{e\} \in \mathcal{S}$}
	\RETURN $\{\{ e \}\} \cup  \mathrm{MSC}(\mathcal{S} \backslash \{ \{e\} \}$)
\ELSIF{$\exists S: \left| \bigcup \{ S' \backslash S | e \in S, f(e)=2, S' \in \mathcal{S}(e) \} \right| < \left| \{ e \in S | f(e)=2 \} \right|$}
	\RETURN $S \cup \mathrm{MSC}(\{S' \backslash S | S' \in \mathcal{S} \backslash \{ S \} \})$
\ELSE
	\STATE Let $S := \mathrm{argmax}_{S' \in \mathcal{S}}(|S'|)$
	\STATE $\mathcal{P} = \{ \mathrm{MSC}(\mathcal{S} \backslash \{ S \}), S \cup \mathrm{MSC}(\{S' \backslash S | S' \in \mathcal{S} \backslash \{ S \} \}) \}$
	\RETURN $\mathrm{argmin}_{P \in \mathcal{P}}(|P|)$
\ENDIF
\STATE \}
\end{algorithmic}
\caption{Algorithm Designed by Measure and Conquer}
\label{algorithm}
\end{algorithm}

\subsection{Running time analysis by Measure and Conquer} \label{sec:m&c}
The basic idea of {\em measure and conquer} is the usage of a non-standard measure 
for the complexity of a problem instance in combination with an extensive subcase analysis.
In the case of {\sc Set Cover}, we give weights to set sizes and element frequencies,
and sum these weights over all items and sets. We enumerate many subcases in which the algorithm
can branch and derive recurrence relations for each of these cases in terms of these weights.
Finally we obtain a large numerical optimisation problem which computes the weights corresponding
to the smallest solution to all recurrence relations, giving an upper bound on the running time of our algorithm.
This analysis is similar to \cite{FominGK05a}.

We let $v_i, w_i \in [0,1]$ be the weight of an element of frequency $i$ and a set of size $i$ respectively,
and set our {\em variable measure of complexity} $k_\mathcal{S}$ to:
\[ k_\mathcal{S} = \sum_{S \in \mathcal{S}} w_{|S|} + \sum_{e \in \mathcal{U_S}} v_{f(e)} 
   \qquad \mathrm{notice:} \quad k_\mathcal{S} \leq d_\mathcal{S} \]
Sets of different sizes and elements of different frequencies contribute equally to the
dimension of the instance, but now larger sets and higher frequency elements can contribute
more to the measured complexity of the instance.
Furthermore we set $v_i = w_i = 0, i \in \{0,1\}$ since all frequency one elements
and size one sets are removed by the reduction rules.
For later use we introduce quantities representing the reduction in problem complexity
when the size of a set or the frequency of an element is reduced by one.
For technical reasons, these quantities must be non-negative.
\[ \Delta w_i = w_i - w_{i-1} \qquad \Delta v_i = v_i - v_{i-1} \qquad \forall i \geq 1 : \Delta v_i, \Delta w_i \geq 0 \]

The next step will be to derive recurrence relations representing problem instances the algorithm branches on.
Let $N(k)$ be the number of subproblems generated in order to solve a problem of measured complexity $k$.
And let $\Delta k_{in}$ (include $S$) and $\Delta k_{out}$ (discard $S$) be the difference in measured complexity
of both subproblems compared to the problem instance we branch on.
Finally let $|S| = \sum_{i=2}^\infty r_i$, where $r_i$ the number of elements in $S$ of frequency~$i$.

If we add $S$ to the set cover, $S$ is removed together with all its elements.
This results in a reduction in size of $w_{|S|} + \sum_{i=2}^\infty r_i v_i$.
Because of the removal of these elements, other sets are reduced in size;
this leads to an additional complexity reduction of at least $\min_{j \leq |S|} \{ \Delta w_j \} \sum_{i=2}^\infty (i-1)r_i$.
To keep the formula (and the next) linear, we set $\min_{j \leq |S|} \{ \Delta w_j \} = \Delta w_{|S|}$
and keep the formula correctly modelling the algorithm by introducing the following constraints on the weights:
\[ \forall i \geq 2 : \Delta w_i \leq \Delta w_{i-1} \]
One can show that including these in the numerical optimisation problem
does not change the solution, as it gives the same weights.
The constraints help to considerably speed up this optimisation process.

If we discard $S$, we also remove it from the problem instance, and hence all its elements are reduced in frequency by one.
So we have a complexity reduction of $w_{|S|} + \sum_{i=2}^\infty r_i \Delta v_i$.
Besides this reduction, the sets which contain the second occurrences of any frequency two element
are included in the set cover. Notice that these must be different sets due to reduction Rule \ref{rule:subsumption}.
Because of Rule \ref{rule:special} we know that at least $r_2$ other elements must be in these sets as well,
and these also must occur somewhere else in the instance, hence even more sets are reduced in size.
Summation leads to an additional size reduction of $r_2(v_2 + w_2 + \Delta w_{|S|})$.
Here we also use Rule \ref{rule:component} to make sure that not all these frequency two elements
occur in the same set, because in that case all considered sets form a connected component of at most five sets
which thus can be solved in $O(1)$ time.

This leads to the following set of recurrence relations: $\forall \; 3 \leq |S|=\sum_{i=2}^\infty r_i$:
\[ N(k) \leq  N(k - \Delta k_{out}) + N(k - \Delta k_{in}) \]
where
\begin{eqnarray}
\Delta k_{out} & = &  w_{|S|} + \sum_{i=2}^\infty r_i \Delta v_i + r_2(v_2 + w_2 + \Delta w_{|S|})   \nonumber \\
\Delta k_{in}  & = &  w_{|S|} + \sum_{i=2}^\infty r_i v_i + \Delta w_{|S|} \sum_{i=2}^\infty (i-1) r_i    \nonumber
\end{eqnarray}

We make the problem finite by setting for some large enough $p$ all $v_i = w_i = 1$ for $i \geq p$,
and only consider the subcases $|S|=\sum_{i=2}^{p} r_i + r_{>p}$, where $r_{>p} = \sum_{i=p+1}^\infty r_i$.
Now we have a finite set of recurrences which model our algorithm since
the recurrences for the cases where $|S| > p+1$ are dominated by those where $|S| = p+1$.
The best value for $p$ follows from the optimisation, for if chosen too small the now constant
recurrences (weights equal~1) will dominate all others in the optimum, and if chosen too large
the extra $v_i$ and $w_i$ are optimised to 1 (and the optimisation problem was unnecessarily hard).
Here $p$ equals~$7$.

A solution to this set of recurrence relations will be of the form $N(k) = \alpha^k$, 
where $\alpha$ is the smallest solution of the set of inequalities:
\[ \alpha^k \leq \alpha^{k-\Delta k_{out}} + \alpha^{k-\Delta k_{in}} \]
Since $k \leq d$ where $d$ the dimension of the problem, we know that the algorithm
will have a running time of $O((\alpha+\epsilon)^d)$, for any $\epsilon > 0$:
\[ O(poly(d)N(k)) = O(poly(d) \alpha^k) \leq O(poly(d) \alpha^d) \leq O((\alpha + \epsilon )^d) \]
From here on we let $\epsilon$ be the error in the upward decimal rounding of $\alpha$.

So for any given vector $\vec{v} = (0, v_2, v_3, v_4, \ldots)$ and $\vec{w} = (0, w_2, w_3, w_4, \ldots)$
we can now compute the running time measured with these weights.
As a result we have obtained a numerical optimisation problem: choose
the best weights so that the upper bound on the running time is minimal.

The numerical solution to this problem can be found in the last cell of Table \ref{tab:steps},
resulting in an upper bound on the running time of the algorithm of $O(1.2302^d)$:

\subsection{Quasiconvex programming}
The sort of numerical optimisation problems arising from
measure and conquer analyses are {\em quasiconvex programs}, named after
the kind of function we are optimising: a {\em quasiconvex function},
which is a function with convex level sets $\{ \vec{x} \:|\: q(x)\leq \lambda \}$.

To our knowledge there are currently two different techniques in use to solve
these quasiconvex programs: randomised search, and  Eppstein's \emph{smooth
quasiconvex programming} algorithm \cite{Eppstein04}.
We have implemented a variant of the second technique; for details see \cite{vanRooij06}.

\subsection{Results}
As discussed, we have now obtained the following result.
\begin{theorem}
Algorithm \ref{algorithm} solves a {\sc Set Cover}
problem instance of dimension $d$ in $O(1.2302^d)$ time and polynomial space.
\qed
\end{theorem}
Using the minimum set cover modelling of {\sc Dominating Set} this results in:
\begin{corollary}
There exists an algorithm that solves the {\sc Dominating Set} problem in $O(1.5134^n)$ time and polynomial space.
\qed
\end{corollary}

We can further reduce the time complexity of the algorithm at the cost of exponential space.
This can be done by dynamic programming; the algorithm keeps track of all solutions to
subproblems solved and if the same subproblem turns up more than ones it is looked up.
Notice that querying and storing the subproblems can be implemented in polynomial time.

We compute the new time complexity based on \cite{FominGK07,Robson86} and obtain:
\begin{theorem} \label{thrm:exptime}
Algorithm \ref{algorithm}, modified as above, solves a {\sc Set Cover}
problem of dimension $d$ in $O(1.2273^d)$ time and space.
\qed
\end{theorem}
\begin{corollary}
There exists an algorithm that solves the {\sc Dominating Set} problem in $O(1.5063^n)$ time and space.
\qed
\end{corollary}

\section{Design by Measure and Conquer}
\label{section:design}
The beauty of our algorithm lies in the fact that it has been designed
using a form of {\em computer aided algorithm design} which we call {\em design by measure and conquer}.
Given a variable measure of complexity as in the analysis in Section \ref{sec:m&c}
and a set of branching rules, all polynomial time computable reduction rules
relative to this measure and branching rules follow by the method.
We start with a trivial branch and reduce algorithm, i.e. one without any reduction rules and only
consisting of the branching rule and a trivial base case (if the problem is empty, return $\emptyset$).
Next we exhaustively apply an improvement step,
which comes up with a new reduction rule and hence a possibly faster algorithm.
This changes the algorithm analysis technique measure and conquer into a technique for algorithm design.

Thus, this gives a very nice process, where a human invents 
additional reduction rules, and the computational power of our computer does
the extensive measure and conquer analysis and
points to all possible points of direct improvement.
This combination has proven to be successful
as we see from the results of Section \ref{sec:m&calgorithm}.
While constructing our algorithm, the previously fastest algorithm for {\sc Dominating Set}
by Fomin et al. \cite{FominGK05a} has been obtained as an intermediate step.
It has now been improved up to a point where we need to either
change the branching rule (or add new branching rules) or modify the measure and conquer analysis,
i.e. use a different variable measure or perform a more elaborate subcase analysis.
See Table \ref{tab:steps} for information on the analysis and added rules for
all algorithms, from the starting trivial algorithm without any reduction
rules till we obtain Algorithm \ref{algorithm}.

\subsection{A Single Iteration: improving the previously fastest algorithm}
We now demonstrate how the improvement step works, by giving one such
improvement as elaborate example, namely an improvement we can make when we
start with the algorithm by Fomin et al.~\cite{FominGK05a}.
This step is marked with a star in Table \ref{tab:steps}.
First we perform a measure and conquer analysis on the current algorithm
giving us the optimal instantiation of the variable measure,
and an upper bound on the running time of the algorithm.
Next we examine the quasiconvex function we have just optimised.

The quasiconvex function has the following form:
\[ q(\vec{v}, \vec{w}) = \max_{c \in \mathcal{C}} q_c(\vec{v}, \vec{w})= \max_{c \in \mathcal{C}} \left\{ \alpha \in \mathbb{R}_{>0} \; | 1 = \alpha^{-\Delta k_{out}^c} + \alpha^{-\Delta k_{in}^c} \right\} \]
where $\mathcal{C}$ is the set of all possible instances the algorithm can branch on
and $\Delta k_{in}^c, \Delta k_{out}^c$ are the differences in measured complexity between
the generated subproblems and branching subcase $c$.

Each one of the functions $q_c$ is quasiconvex (see \cite{Eppstein04}), i.e. it has convex level sets.
The situation is very similar to finding the point $x$ of minimum maximum distance to
a set of points $P$ in $N$ dimensional space:
only a few points in $P$ have distance to $x$ tight to this maximum,
and moving away from $x$ always results in at least one of these distances to increase.
If one such tight point is moved or removed, this directly influences the optimum $x$ and the minimum maximum distance.


We now consider the eight cases that are tight to the value of the 
quasiconvex function $q$ in the optimum.
These are:
\begin{eqnarray}
|S|=r_2=3, \quad |S|=r_3=3, \quad |S|=r_4=3, \quad |S|=r_5=4 \nonumber \\
|S|=r_6=4, \quad |S|=r_6=5, \quad |S|=r_7=5, \quad |S|=r_7=6 \nonumber
\end{eqnarray}
Now, if we can formulate a reduction rule that either further reduces the size of any subproblem generated in these cases,
or removes any of these cases completely, then we lower the value of the corresponding $q_c$,
or remove this $q_c$ respectively, resulting in a new optimum corresponding to a faster running time.

We take the simplest case for improvement; $|S|$ as small as possible,
and with as low frequency elements as possible.
This corresponds to an instance with:
\[ S = \{ \, 1,2,3 \: \} \quad \textrm{existing next to:} \quad \{ \, 1,2,4 \: \}, \{ \, 3,4 \: \}\]
We emphasize that this is not an entire instance, but just a fragment of an instance containing the set $S$ used for branching 
and the collection of sets in which the elements from $S$ also occur.
In an instance corresponding to this subcase the element 4 can be of frequency two or higher, but all sets are of size three or smaller.

We note that we do not need to branch on this particular subcase:
elements 1 and 2 occur in exactly the same sets, and thus
if a set cover covers one of these, the other is covered as well.
We generalise this and formulate the subsumption rule (Rule
\ref{rule:subsumption} of Algorithm \ref{algorithm}).
Now we have a new algorithm, for which we can adjust the measure and conquer analysis, and repeat this process.

\begin{table} \begin{center}
\begin{tabular}{|ll|}
\hline
\textbf{Latest new reduction rule}		& Running times for {\sc Set Cover} and \\
current formula for $\Delta k_{out}$	& \hspace*{5pt} {\sc Dominating Set} \\
subcases considered 					& instance part of the simplest worst case; \\
weights vectors $\vec{v}$ and $\vec{w}$	& \hspace*{5pt} $S \; - $ other occurrences of elements of $S$\\
\hline
\multicolumn{2}{l}{}\\
\hline
\textbf{Trivial algorithm}						& $O(1.4519^d) \quad O(2.1080^n)$ \\
\multicolumn{2}{|l|}{ $w_{|S|} + \sum_{i=1}^\infty r_i \Delta v_i$ } \\
$1 \leq |S| = \sum_{i=1}^p r_i + r_{>p} \leq p + 1 = 3$
& $\{ \, 1 \: \} \; - \; \emptyset$ \\
$\vec{v} = (0.8808, 0.9901, \ldots )$ & $\vec{w} = (0.9782, \ldots )$ \\
\hline
\textbf{Stop when all sets of size one}			& $O(1.3380^d) \quad O(1.7902^n)$ \\
\multicolumn{2}{|l|}{ $w_{|S|} + \sum_{i=1}^\infty r_i \Delta v_i$ }\\
$2 \leq |S| = \sum_{i=1}^p r_i + r_{>p} \leq p + 1 = 4$
& $\{ \, 1, 2 \: \} \; - \; \emptyset$ \\
$\vec{v} = (0.7289, 0.9638, 0.9964, \ldots )$ & $\vec{w} = (0.4615, 0.9229, \ldots )$ \\
\hline
\textbf{Include all frequency one elements}		& $O(1.2978^d) \quad O(1.6842^n)$ \\
\multicolumn{2}{|l|}{ $w_{|S|} + \sum_{i=2}^\infty r_i \Delta v_i + \delta_{r_2>0}w_1 + \delta_{|S|=r_2=2}\Delta w_2$ } \\
$2 \leq |S| = \sum_{i=2}^p r_i + r_{>p} \leq p + 1 = 5$
& $\{ \, 1,2 \: \} \; - \; \{ \, 1,2 \: \}$ \\
$\vec{v} = (0, 0.4818, 0.8357, 0.9636, \ldots )$ & $\vec{w} = (0.4240, 0.8480, 0.9676, \ldots )$ \\
\hline
\textbf{Subset rule}							& $O(1.2665^d) \quad O(1.6038^n)$ \\
\multicolumn{2}{|l|}{ $w_{|S|} + \sum_{i=2}^\infty r_i \Delta v_i + \delta_{r_2>0}(w_2+v_2) + \delta_{|S|=r_2=2}\Delta w_2$ } \\
$2 \leq |S| = \sum_{i=2}^p r_i + r_{>p} \leq p + 1 = 7$
& $\{ \, 1,2 \: \} \; - \; \{ \, 1,3 \: \}, \{ \, 1,4 \: \}, \{ \, 2,3 \: \}, \{ \, 2,4 \: \}$\\
$\vec{v} = (0, 0.3900, 0.7992, 0.9318, 0.9808, \ldots)$ & $\vec{w} = (0, 0.6973, 0.9093, 0.9800, \ldots)$ \\
\hline
\textbf{Compute matching for size two sets$^*$} & $O(1.2352^d) \quad O(1.5258^n)$ \\
\multicolumn{2}{|l|}{ $w_{|S|} + \sum_{i=2}^\infty r_i \Delta v_i + \delta_{r_2>0}(w_2+v_2) + \delta_{|S|=3, r_2 \geq 2}(\Delta w_3 + \delta_{r_2=3}w_2) + \delta_{|S|=r_2=4}w_4$ } \\
$3 \leq |S| = \sum_{i=2}^p r_i + r_{>p} \leq p + 1 = 7$
& $\{ \, 1,2,3 \: \} \; - \; \{ \, 1,2,4 \: \}, \{ \, 3,4 \: \}$ \\
$\vec{v} = (0, 0.3978, 0.7650, 0.9263, 0.9842, \ldots)$ & $\vec{w} = (0, 0.3787, 0.7575, 0.9103, 0.9763, \ldots)$ \\
\hline
\textbf{Subsumption rule}						& $O(1.2339^d) \quad O(1.5223^n)$ \\
\multicolumn{2}{|l|}{ $w_{|S|} + \sum_{i=2}^\infty r_i \Delta v_i + \delta_{r_2>0}(r_2w_2 + v_2) + \delta_{|S|=r_2=3}\Delta v_3$ } \\
$3 \leq |S| = \sum_{i=2}^p r_i + r_{>p} \leq p + 1 = 7$
& $\{ \, 1,2,3 \: \} \; - \; \{ \, 1,4 \: \}, \{ \, 2,4 \: \}, \{ \, 3,4 \: \}$ \\
$\vec{v} = (0, 0.3545, 0.7455, 0.9203, 0.9818, \ldots)$ & $\vec{w} = (0, 0.3755, 0.7510, 0.9061, 0.9745, \ldots)$ \\
\hline
\textbf{Avoid unnecessary branchings}	& $O(1.2313^d) \quad O(1.5160^n)$ \\
\multicolumn{2}{|l|}{ $w_{|S|} + \sum_{i=2}^\infty r_i \Delta v_i + r_2(w_2 + v_2) + \delta_{r_2>1}(r_2-1)\Delta w_{|S|}$ } \\
$3 \leq |S| = \sum_{i=2}^p r_i + r_{>p} \leq p + 1 = 8$
& $\{ \, 1,2,3 \: \} \; - \; \{ \, 1,4 \: \}, \{ \, 2,5 \: \}, \{ \, 3,6 \: \}$ \\
\multicolumn{2}{|l|}{ $\vec{v} = (0, 0.269912, 0.689810, 0.892666, 0.965849, 0.992140, \ldots )$ } \\
\multicolumn{2}{|l|}{ $\vec{w} = (0, 0.376088, 0.752176, 0.907558, 0.974394, 0.999212, \ldots )$ } \\
\hline
\textbf{Connected components (final)}	& $O(1.2302^d) \quad O(1.5134^n)$ \\
\multicolumn{2}{|l|}{ $w_{|S|} + \sum_{i=2}^\infty r_i \Delta v_i + r_2(w_2 + v_2 + \Delta w_{|S|})$ } \\
$3 \leq |S| = \sum_{i=2}^p r_i + r_{>p} \leq p + 1 = 8$
& $\{ \, 1,2,3 \: \} \; - \; \{ \, 1,4 \: \}, \{ \, 2,5 \: \}, \{ \, 3,6 \: \}$ \\
\multicolumn{2}{|l|}{ $\vec{v} = (0, 0.219478, 0.671386, 0.876555, 0.956850, 0.988195, \ldots )$ } \\
\multicolumn{2}{|l|}{ $\vec{w} = (0, 0.375418, 0.750835, 0.905768, 0.971965, 0.998158, \ldots )$ } \\
\hline
\multicolumn{2}{l}{ $^*$ Algorithm by Fomin, Grandoni and Kratsch \cite{FominGK05a}.}
\end{tabular} \end{center}
\caption{The iterations of the design by measure and conquer process.}
\label{tab:steps}
\end{table}

\subsection{The Process Halts}
Above, we discussed 
how to perform one step of the design by measure and conquer process.
For a complete overview of the construction of Algorithm \ref{algorithm} see
Table \ref{tab:steps}, with the relevant data for each improvement step.
Note from Table \ref{tab:steps}
that after each new step, the example worst case instance part
is no longer a valid worst case for the next step.
As a result, at each step either some subcases are removed by using
a larger smallest set $S$ or by removing small sets or elements (setting $v_1=0$ or $w_1=0$),
or the size reduction in the formula for $\Delta k_{out}$ is increased.
After each step we refactored the reduction rules and removed possible superfluous ones.
We have not included the formula for $\Delta k_{in}$ in this table,
since it does not change except that $r_1 \!\not= 0$ in early stages.

It appears that we must use a different approach to obtain a faster algorithm. 
Considered the following problem:
\begin{problem} \label{NPCproblem}
Given a {\sc Set Cover} instance $\mathcal{S}$ and a set $S \in \mathcal{S}$ with the properties:
\begin{enumerate}
\item Non of the reduction rules of Algorithm \ref{algorithm} apply to $\mathcal{S}$.
\item All sets in $\mathcal{S}$ have cardinality at most three; $|S|=3$.
\item Every element $e \in S$ has frequency two.
\end{enumerate}
Question: Does there exist a minimum set cover of $\mathcal{S}$ containing $S$?
\end{problem}

\begin{proposition} \label{prop:npc}
Problem \ref{NPCproblem} is NP-complete.
\end{proposition}

Proposition \ref{prop:npc} implies that we cannot formulate a polynomial time reduction rule that removes the
current simplest worst case of our algorithm by deciding on whether $S$ is in a minimum set cover or not, unless $P=NP$.

We can construct similar NP-complete problems for all other worst cases of Algorithm \ref{algorithm}.
Therefore Algorithm \ref{algorithm} is optimal in some sense:
we cannot straightforwardly improve it by performing another iteration.
In order to obtain a faster branch and reduce algorithm using polynomial time reduction rules 
with a smaller measure and conquer proved time bound,
it is necessary to either change the variable measure, the branching rule(s),
or perform a more extensive subcase analysis.

Very recently, we pursued the last option with little result.
We tried to further subdivide the frequency two elements in the branch set
depending on the size of the set containing their second occurence (two or larger) and
if this is a set of size two, on the frequency of the other element in this set.
This resulted in a set of very technical reduction rules
and a small speedup for the case where we use only polynomial space.
This speedup, however, was almost completely lost when using exponential space
because some of the weights involved became almost zero.

\section{Conclusion and Further Research}
In this paper, we have given the currently fastest exact algorithm for the {\sc Dominating Set}
problem. Besides setting the current record for this central graph theoretic problem, we also
have shown that measure and conquer can be used as a tool for the design of algorithms.

We have shown that there exists a strong relation between the chosen variable measure,
the branching rule(s) and the reduction rules of a measure and conquer based algorithm.
We intend to further investigate this relation and 
examine to what point we can deduce not only reduction rules,
but also branching rules from the given measure.

We plan to apply the {\em design by measure and conquer} method to a number of other combinatorial
problems, and hope and expect that in a number of cases, such a computer aided algorithm design
will give further improvements to the best known exact algorithms for these problems.

In this paper, we observe that measure and conquer can be used as a form of
{\em computer aided algorithm design}.
Another intriguing question is whether we can automate some additional steps in the
design process, e.g., can we automatically obtain reduction rules from the 
solution of the quasiconvex program?

\newpage
\null

\end{document}